\documentclass{aa} % for a referee version
\usepackage{graphicx}
\usepackage{txfonts}
\begin{document}
   \title{Formation of giant globular cluster G1 and the origin of the M31 stellar halo}

%%   \subtitle{I. Overviewing the $\kappa$-mechanism}

   \author{K. Bekki
          \inst{1}
          \and
          M. Chiba \inst{2} 
%%%\fnmsep\thanks{Just to show the usage
%%%          of the elements in the author field}
          }

   \offprints{K. Bekki}

   \institute{School of Physics, University of New South Wales, Sydney 2052, Australia\\
              \email{bekki@bat.phys.unsw.edu.au}
         \and
             Astronomical Institute, Tohoku University, Sendai, 980-8578, Japan\\
              \email{chiba@astr.tohoku.ac.jp}
%%%\thanks{The university of heaven temporarily does not
%%%                     accept e-mails}
             }

   \date{Received 20 September 2003/ Accepted 21 Octorber 2003}

   \abstract{
We first demonstrate that globular cluster G1 could have been formed
by tidal interaction between M31 and a nucleated dwarf galaxy (dE,N).
Our fully self-consistent numerical simulations 
show that during tidal interaction between M31 and G1's progenitor  dE,N 
with $M_{\rm B}$ $\sim$ $-15$ mag and its nucleus mass of $\sim$ $10^7$ $M_{\odot}$,
the dark matter and the outer stellar envelope of the dE,N are nearly
completely stripped whereas the nucleus can survive  the tidal stripping
because of its initially compact nature.
The naked nucleus (i.e., G1) has orbital properties similar to those
of its progenitor dE,N. 
The stripped stars form a metal-poor ([Fe/H] $\sim$ $-1$) stellar halo around M31
and its structure and kinematics depend strongly on the initial orbit of
 G1's progenitor dE,N. 
We suggest that the observed large projected distance of G1 from M31 ($\sim$ 40 kpc) 
can give some strong constraints
on the central density of the dark matter halo of  dE,N.
We discuss these results in the context 
of substructures  of M31's stellar halo recently revealed  by
Ferguson et al. (2002).
   }

   \authorrunning{K. Bekki and M. Chiba}
   \titlerunning{Formation of G1}
   \maketitle
%
%________________________________________________________________

\section{Introduction}

A growing number of photometric and spectroscopic observations
have suggested  that 
G1 (=Mayall II),  
which is one of the brightest globular clusters
belonging to M31,  has very unique physical properties 
as  a globular cluster (e.g., Meylan et al. 2001).
These include a possible intrinsic metallicity dispersion
among its stellar population (Meylan et al. 2001),
the large central velocity dispersion of $\sim$ 25 km s$^{-1}$
(e.g., Djorgovski  et al. 1997),
the very flattened shape with mean ellipticity of 0.2
and significantly high central surface brightness 
(Rich et al. 1996; Meylan et al. 2001),
and the possible existence of a black hole with a mass of
20000 $M_{\odot}$ (Gebhardt et al. 2002). 
In spite of its  extraordinary nature,
G1 is observed to be on the $M_{\rm V}$-${\sigma}_{0}$ relation
(where $M_{\rm V}$ and ${\sigma}_{0}$ are total magnitude in $V-$band 
and central velocity dispersion)
defined not by elliptical and dwarf galaxies
but by globular clusters, 
which implies  that G1 looks like a genuine globular cluster
(Meylan et al. 2001).

One of the possible scenarios of  G1 formation
is that G1 is the surviving nucleus of an ancient nucleated dwarf galaxy
with its outer  stellar envelope almost entirely stripped by M31's strong
tidal field 
(Meylan et al. 1997, 2000,  2001).
Such a scenario has already been suggested by Zinnecker et al. (1988) and Freeman (1993),
and the viability of the scenario has been extensively discussed by many authors
in terms of $\omega$ Cen formation
(e.g., Hilker \& Richtler 2000; 
Dinescu 2002; Gnedin et al. 2002; Zhao 2002; Bekki \& Freeman 2003; Mizutani et al. 2003).
However, no theoretical attempts have been  made so far to 
investigate  (1) whether M31's tidal filed is strong enough to transform
a dE,N into G1 
and (2) what  observable evidence of the past destruction of
 G1's progenitor dE,N we can find 
in the M31 halo regions. 
The above point (2) is very important,
because Ibata  et al. (2001) and Ferguson et al. (2002)
have recently discovered M31's stellar halo substructures,
which could have been formed from tidal destruction of
M31's satellite dwarfs.

In this paper, by using numerical simulations,
we first demonstrate that G1 can be formed from a dE,N 
during tidal interaction between the dE,N and M31.
Our fully self-consistent
numerical simulations  demonstrate  that the stellar envelope of 
dE,N with $M_{\rm B}$ $\sim$ $-15$
can be nearly completely stripped by the  strong tidal field of M31
whereas the central nucleus can remain intact owing to its compactness.
We suggest that this naked nucleus orbiting M31 is  a giant
globular cluster (i.e., G1).
The morphological transformation from dE,Ns into very compact stellar systems 
was originally  investigated by Bekki et al. (2001, 2003) 
for ultra-compact dwarfs labeled as ``UCD'' (Drinkwater et al. 2003)
and called ``galaxy threshing'' (Bekki et al. 2001), 
though they  suggested that galaxy threshing is also important
for the formation of giant globular clusters such as $\omega$ Cen and G1. 
The present study confirms this early suggestion
and discusses the  physical relationship between UCDs, $\omega$ Cen, and G1.  
%__________________________________________________________________

\section{Model}

%                                     Two column figure (place early!)
%______________________________________________ Gamma_1 (lg rho, lg e)
   \begin{figure*}
   \centering
   \caption{ {\it Upper four}:
Morphological evolution of the stellar envelope 
of the dE,N projected onto the $x$-$z$ plane (i.e., edge-on view) for the
fiducial model. For comparison, the M31 disk and bulge components
are shown 
in the upper left panel and
the disk size is indicated by a solid line in other three panels.
The orbit of the dE,N with $R_{\rm apo}$ = 80 kpc and $e_{\rm}$ = 0.62
is indicated by a dotted  line in the upper left panel.
The time $T$ (in units of Gyr) indicated in the
upper left corner of each frame represents the time  elapsed since
the simulation starts. Each frame is 164\,kpc on a side.
{\it Lower four}: 
Morphological evolution of the stellar envelope 
and the nucleus  of the dE,N projected onto the $x$-$y$ plane (i.e., face-on view) for the
fiducial model. 
Each frame is 20\,kpc on a side.
               }
              \label{Fig.1}%
    \end{figure*}

   \begin{figure*}
   \centering
   \includegraphics{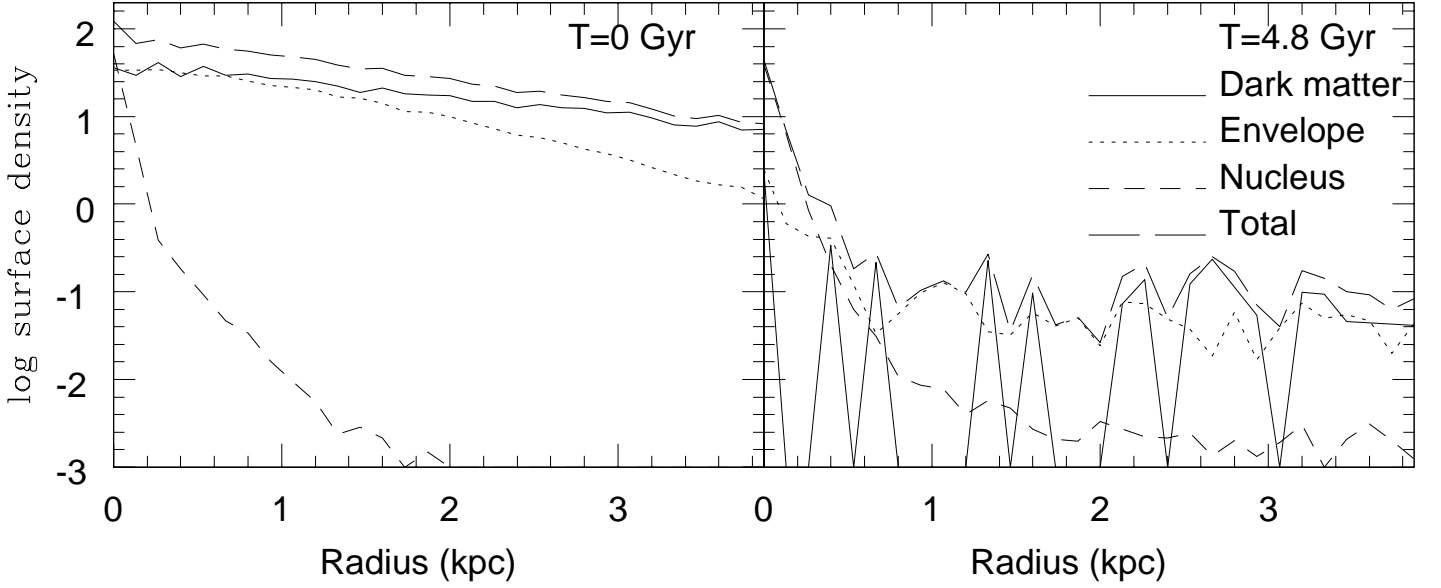}
   \caption{ 
The projected surface density profiles for
the dark matter ({\it solid} line), the stellar envelope ({\it dotted}
line), the nucleus ({\it short-dashed} line), and all these components
({\it long-dashed} line) at $T = 0$\,Gyr ({\it upper} panel) and 4.8\,Gyr
({\it lower} panel) in the fiducial model.
               }
              \label{Fig.2}%
    \end{figure*}

Since our numerical methods and techniques for modeling 
collisionless,  self-gravitating systems
of dE,Ns already have been  described in detail by Bekki et al. (2003),
we give only a brief review here.
We consider that a dE,N with a mass and size similar to
that observed for the dE,N types orbiting  M31. 
The dE,N is modeled as a fully self-gravitating system
and is assumed to consist of a dark matter halo, a stellar component and
a nucleus: We first investigate this ``three-component'' dE,N model. 
For convenience, the stellar component
(i.e., the main baryonic component) is referred to as either
the ``envelope'' or the ``stellar envelope'' so that
we can distinguish this component from the stellar nucleus.
The density profile of the dark matter halo with the total mass of $M_{\rm dm}$
in  the dE,N  is represented by that proposed by
Salucci \& Burkert (2000):
\begin{equation}
{\rho}_{\rm dm}(r)=\frac{\rho_{dm,0}}{(r+a_{\rm dm})(r^2+a_{\rm dm})^2},
\end{equation}
where $\rho_{dm,0}$ and $a_{\rm dm}$ are the central dark matter
density and the core (scale) radius, respectively.
This model profile is consistent with observations
and different from the predictions of the standard CDM model
(Navarro, Frenk, \& White 1996, hereafter NFW).
The dark matter core parameters, $\rho_{dm,0}$,  $a_{\rm dm}$,  and $M_{0}$
(where $M_{0}$ is the total dark matter mass within $a_{\rm dm}$)
have a  clear observed  correlation,
$M_{0}=4.3 \times 10^7 {(\frac{a_{\rm dm}}{\rm kpc})}^{7/3} M_{\odot}$
(Burkert 1995). 
%Therefore, if we give $M_{\rm dm}$ (or $M_{\rm 0}$),
%$\rho_{dm,0}$ and $a_{\rm dm}$ are are automatically determined. 

The mass (luminosity) and the scale length of the stellar envelope of the  dE,N
is modeled according to the observed scaling relation of Ferguson \&
Binggeli (1994):
\begin{equation}
{\rm log}a_{\rm dw} [{\rm pc}] = -0.02 M_{\rm B} +2.6
\end{equation}
for faint dwarfs ($M_{\rm B}$ $\ge$ $-16$), where
$a_{\rm dw}$ and $M_{\rm B}$ are the scale length of the exponential
profile and the absolute $B-$band magnitude, respectively.
The projected density of the envelope with $M_{\rm B}$ and the total mass
of $M_{\rm dw}$ (and $M_{\rm dw}/L_{\rm B}$ = 2)
is represented by an exponential profile with a scale
length $a_{\rm dw}$. 
The projected density profile of the nucleus with mass $M_{\rm n}$
is represented by a King model (King 1964) with a core radius of
$a_{\rm n}$ and a central concentration parameter $c$ of 1.0.

The nuclei typically contribute about a few percent  of
the total light of dwarfs (Binggeli \& Cameron 1991; Freeman 1993)
and the present-day  mass of G1 is 
estimated to be 1.5 $\times$ $10^7$ $M_{\odot}$ for the King model (Meylan et al. 2001).  
Considering these observations,
the reasonable  $M_{\rm B}$ of  dE,N
is estimated to be $\sim$ $-15$ mag. 
Given a value of  $M_{\rm dm}$/$M_{\rm dw}$, 
we can determine $a_{\rm dm}$ from the above $M_{\rm 0}$-$a_{\rm dm}$  relation
(Burkert 1995).
For convenience, $a_{\rm dm}$ for $M_{\rm B}$ = $-15$ mag and $M_{\rm dm}$/$M_{\rm dw}$ = 5 
is referred to as $a_{\rm dm,0}$ hereafter.
We mainly investigate the dE,N models with 
$M_{\rm B}$ = $-15$ mag,  $a_{\rm dw}$ = 790 pc,
$M_{\rm dm}$/$M_{\rm dw}$ = 5, 
$a_{\rm dm}$ = $a_{\rm dm,0}$,
$M_{\rm n}$/$M_{\rm dw}$ = 0.05, and $a_{\rm n}$/$a_{\rm dw}$ = 0.02.  
All of these  values are reasonably consistent  with
observations (Binggeli \& Cameron 1991;  Ferguson \& Binggeli 1994).
We also investigate the models with $a_{\rm dm}$ = $0.25a_{\rm dm,0}$
to clarify the importance of the dark matter halo structure of a dE,N
in the formation processes of G1.

M31 is assumed to have the disk mass of 7.8 $\times$ $10^{10}$ $M_{\odot}$,
and the bulge-to-disk-ratio of  0.25, and the maximum rotational velocity of 
260 km s$^{-1}$, all of which are consistent with observations (e.g., van den Bergh 2000).
The initial  disk plane of M31 is set to be the $x$-$y$ plane of a simulation.
The dE,N orbiting M31 has the initial position 
of ($x$, $y$, $z$) = ($R_{\rm apo} \cos \theta$, 0, $R_{\rm apo} \sin \theta$) 
and the initial velocity of ($v_{\rm x}$, $v_{\rm y}$, $v_{\rm z}$) 
= (0, $\alpha V_{\rm c}$, 0),
where $R_{\rm apo}$, $\theta$, and $V_{\rm c}$ are
the apocenter of the orbit,  the inclination angle 
with respect to  the M31's disk (i.e., the $x$-$y$ plane),
the circular velocity at the apocenter,
and the parameter (0 $\le$ $\alpha$ $\le$ 1) 
that determines the orbital eccentricity represented by $e_{\rm p}$ (i.e,
the larger $\alpha$ is, the more circular  the orbit is).  
G1 is observed to have  the projected distance of  40 kpc from M31 
and radial velocity of $-31$  km s$^{-1}$ with respect to M31
(e.g., Meylan et al. 2001).
Guided by these observations,
we investigate the models with $\theta$ = 30$^{\circ}$ and 80$^{\circ}$,
$R_{\rm apo}$ = 40, 80, and 160  kpc,
and $e_{\rm p}$ = 0.62 ($\alpha$ = 0.5) and  0.18 ($\alpha$ = 0.9).
We however describe the four representative and  important models in this paper
and the parameter values in these models (e.g., ``fiducial model'') are given in Table 1.
All the simulations have been carried out on a
GRAPE board (Sugimoto et al. 1990) with the particle
number of 90000.

%Table original
%\begin{tabular}[pos]{3}
%3 & 3 & 3 \\
%\end{tabular}

%__________________________________________________ One column table
   \begin{table}
      \caption[]{Model parameters}
%         \label{KapSou}
     $$ 
         \begin{array}{p{0.5\linewidth}llll}
            \hline
            \noalign{\smallskip}
            model      &   a_{\mathrm{dm}}/a_{\mathrm{dm,0}} 
           & \theta (degrees)  
           & R_{\mathrm{apo}} (kpc) 
           & e_{\mathrm{p}} \\
            \noalign{\smallskip}
            \hline
            \noalign{\smallskip}
fiducial  & 1.0 &  30 & 80 &  0.62  \\
smaller $e_{\rm p}$   & 1.0 &  30 & 80 &  0.18  \\
smaller $R_{\rm apo}$   & 1.0 &  30 & 40 &  0.62  \\
more compact dark matter  & 0.25 &  30 & 80 &  0.62  \\
            \noalign{\smallskip}
            \hline
         \end{array}
     $$ 
%\begin{list}{}{}
%\item[$^{\mathrm{a}}$] This is footnote a
%\end{list}
   \end{table}

%

%
%                                                One column figure
%----------------------------------------------------------- S_vib
%   \begin{figure}
%   \centering
%   %%%\includegraphics[width=3cm]{empty.eps}
%      \caption{Vibrational stability equation of state
%               $S_{\mathrm{vib}}(\lg e, \lg \rho)$.
%               $>0$ means vibrational stability.
%              }
%         \label{FigVibStab}
%   \end{figure}
%
%______________________________________________________________

   \begin{figure*}
   \centering
   \includegraphics[angle=-90]{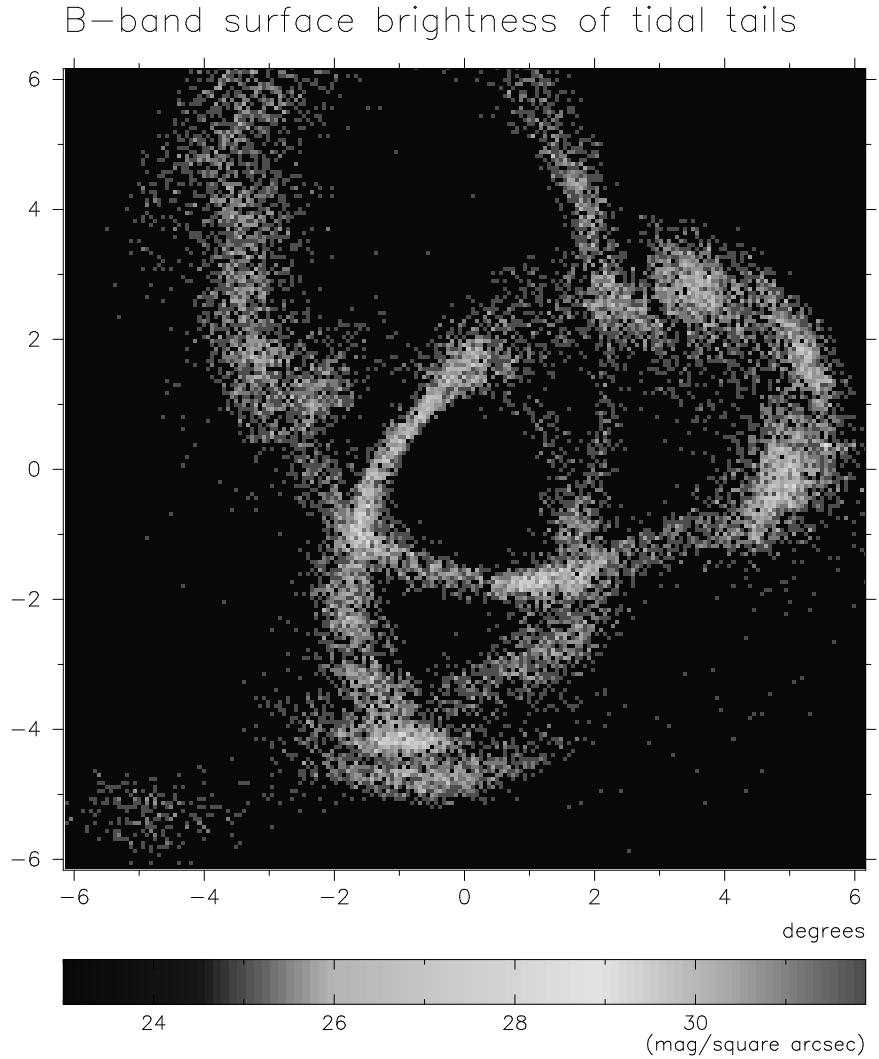}
   \caption{ 
The final $B-$band surface brightness $\mu_{B}$ (mag arcsec$^{-2}$) 
distribution projected onto the $x$-$y$ plane for
substructures/tails of M31's stellar halo developed from tidal destruction of the dE,N
in the fiducial model.
For comparison with observations by Ferguson et al. (2002),
the scale is given in units of degree. 
               }
              \label{Fig.3}%
    \end{figure*}

   \begin{figure*}
   \centering
   \includegraphics{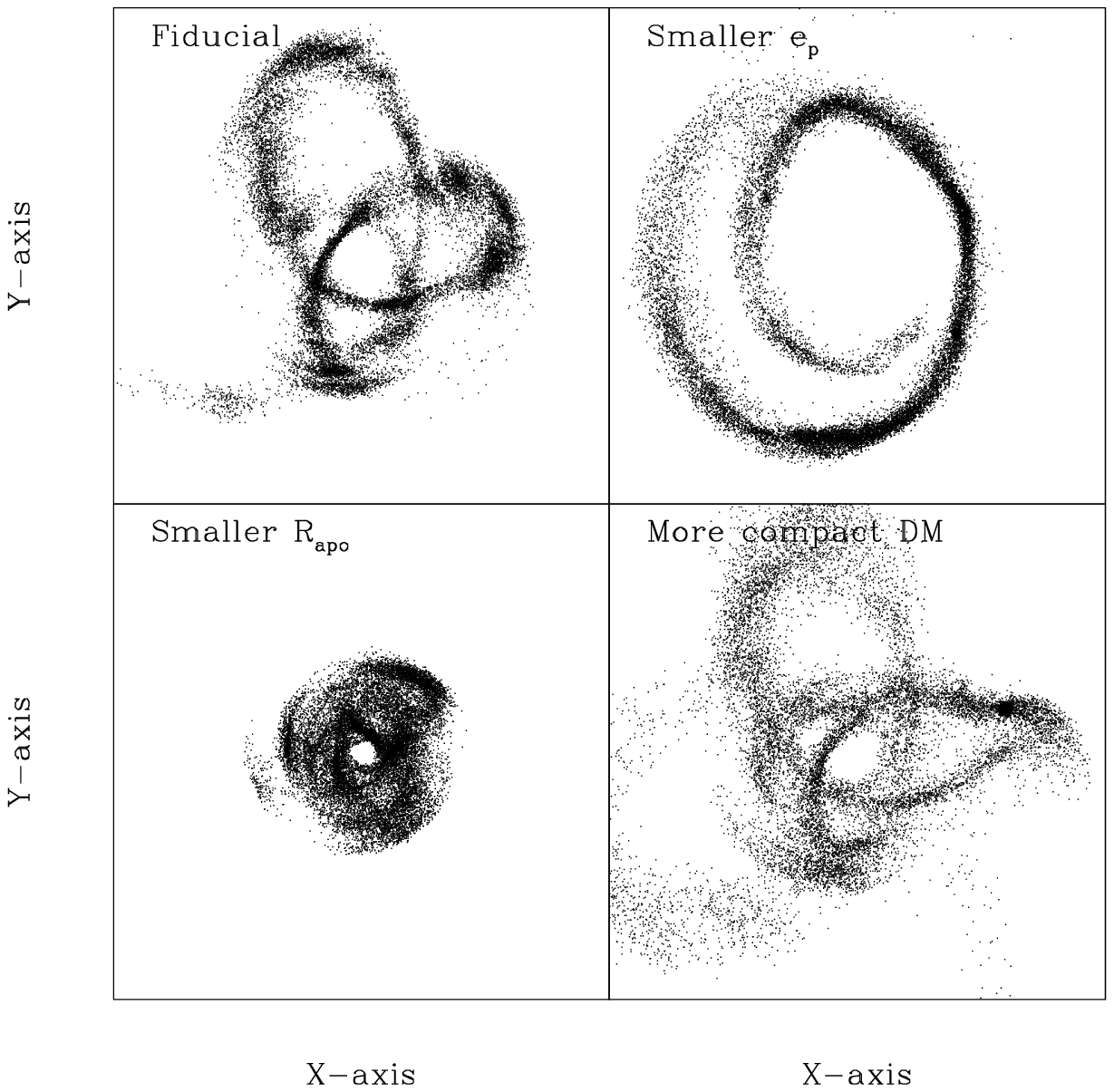}
   \caption{
Final mass distributions projected onto the  $x$-$y$ plane at $T$ = 4.8\,Gyr
for the fiducial model (upper left), 
the smaller $e_{\rm p}$ one  with $e_{\rm p}$ = 0.18 (upper right),
the smaller $R_{\rm apo}$ with $R_{\rm apo}$ = 40 kpc (lower left),
and the more compact dark matter model with $a_{\rm dm}$ = $0.25a_{\rm dm,0}$
(lower right).
Each frame is 240 kpc on a side.
               }
              \label{Fig.4}%
    \end{figure*}

\section{Results}

Figures 1  summarizes the morphological  evolution of the envelope
and the nucleus of the dE,N in the fiducial model,
which shows typical behavior in morphological transformation from
a dE,N into G1.
As the dE,N approaches the pericenter of its orbit,
the strong global tidal field of M31 
stretches the envelope of the dE'N along the direction of the dwarf's orbit 
and consequently tidally strips the stars of the envelope ($T = 0.6$\,Gyr). 
The dark matter halo, which is more widely distributed than the envelope 
due to its larger core radius, is also efficiently removed from the dE,N 
during the pericenter passage. Since the envelope (and the dark matter halo) 
loses a significant fraction of its mass during the passage through the 
pericenter, the envelope becomes more susceptible to the tidal effects of 
M31 after the pericenter passage. Therefore, each subsequent
time the dwarf approaches the pericenter, it loses an increasingly larger
fraction of its stellar envelope through tidal stripping.
Finally, the envelope and the dark matter halo lose 98 \% and 99 \% of
their initial masses, respectively, 
after four passages through
the pericenter  ($T = 4.8$\,Gyr).

The central nucleus, on the other hand, is only  weakly influenced by the
tidal force as a result of its compact configuration. Because of its 
strongly self-gravitating nature, the nucleus loses only a small amount 
($\sim 7$\%) of its mass and thus maintains its compact morphology during 
its tidal interaction with M31. As a result, a very 
compact stellar system with a negligible amount of dark matter
is formed from the dE,N by $T = 4.8$\,Gyr. The total nuclear stellar mass of  
the remnant within $5a_{\rm dw}$ is $\sim$ 1.5 $\times$ $10^7$ $M_{\odot}$, 
consistent with the observed mass  of G1. 
The mass-to-light ratio, $M/L_{\rm B}$, decreases 
dramatically from $\sim$ 10  to $\sim$ 3 for $r < 5a_{\rm dw}$ within 4.8\,Gyr.
This result clearly 
explains why  G1 is  observed to have mass-to-light ratios that are
much smaller  than those observed for dE,Ns ($\sim$ 
10): galaxy threshing is  most efficient in the outer regions of a dE,N 
where the dark matter halo dominates gravitationally. 
As is shown in  Figure 2,
both the surface density of the dark matter and that 
of the envelope drop by more than an order of magnitude within 4.8\,Gyr.
If the initial central surface brightness of the dE,N  is 
$\mu_{B}=23$\,mag\,arcsec$^{-2}$, then the final surface brightness of the dE,N
at $r = 3$\,kpc is about $\mu_{B}=29.5$\,mag\,arcsec$^{-2}$. Such a faint, 
low surface brightness envelope will be hard to detect, even by existing 
large ground-based telescopes.

The stripped stars from the envelope of the G1 progenitor dE,N
form substructures and tidal tails  in  M31 halo region.
Figure 3 shows that the tidal tail is  a ``rosette'',
reflecting the dE,N eccentric orbit ($e_{\rm p}$ = 0.62).
Most of the stellar components of the tail
can be regarded as being
located well outside the M31 inner halo with  $r$ $<$ 40 kpc ($\sim$ three degrees).
The $B-$band surface brightness (${\mu}_{\rm B}$) of the tail ranges  from 
28.7 mag arcsec$^{-2}$ to 30.4 mag arcsec$^{-2}$ 
with a mean of 30.1 mag arcsec$^{-2}$.
These results imply that the tidal streams
formed from the G1 progenitor dE,N  would be  hard
to  detect in previous observations by Ibata et al. (2001) and Ferguson et al (2002)
that mapped the M31 halo region within  $r$ $\sim$ 40 kpc.

Morphological properties of tails and substructures developed
during tidal destruction of a G1 progenitor dE,N depend 
not only on the initial orbit but also on  the dark matter structure of the dE,N.
Figure 4 summarizes the following three dependences.
Firstly, for  the model with a smaller $e_{\rm p}$ (= 0.18),
in which the time scale of morphological transformation from the  dE,N into G1 is
much longer than that of the fiducial model,
the shape of the developed tail is more like a spiral  than a rosette.
Secondly, the distribution of stars in the developed substructures
for the model with a smaller $R_{\rm apo}$ (= 40 kpc) appears to have
a sharp boundary within which the stars can be located in a relatively
homogeneous manner.
Thirdly, the surface density of stars along the tail 
in the model with a more compact dark matter halo 
(with $a_{\rm dm}$ = 0.25 $a_{\rm dm,0}$) is much lower than
that of the fiducial model 
owing to the smaller number  of stripped stars.
Formation of G1 from the dE,N does not occur in this model 
with $a_{\rm dm}$ = 0.25 $a_{\rm dm,0}$
which suggests that a dE,N with the high central density of the dark matter
(such as  seen in  NFW models)  are less likely to be a  G1 progenitor dE,N
(See Bekki et al. 2003 for more discussions on this point).

\section{Discussion and conclusion}

If G1 originates from a dE,N with $M_{\rm B}$ = $-15$ mag,
the mean metallicity of the M31 stellar halo consisting
of stars stripped from the dE,N 
is roughly estimated as [Fe/H] = $-0.96$ (for $B-V$ = 0.71 in the dE,N) 
using the observed metallicity-luminosity relation for dwarf galaxies,
[Fe/H]=$-3.43(\pm 0.14)-0.157(\pm 0.012)\times M_{V}$
(e.g., C\^ot\'e et al. 2000).
Because of its small mass (less than  $10^9$ $M_{\odot}$) and low metallicity,
the stellar halo formed from the tidal destruction of the G1 progenitor dE,N
could not be the major component of the observed high density and metal-rich
([m/H] $\sim$ $-0.5$) M31' stellar halo (Durrell et al. 2000).
Also, relatively metal-rich components ([Fe/H] $\sim$ $-0.7$)
of the stellar tail recently discovered
in the M31 halo region (Ibata et al. 2001)
might be less likely to be formed  from the metal-poor stars stripped from G1's progenitor
dE,N.

Recently Ferguson et al. (2002)  discovered a stellar halo substructure
located in the proximity of G1 (referred to as ``G1 clump'') and showed
that total $V-$band magnitude and $V-$band   surface brightness
of the G1 clump can be estimated to be $-12.6$ mag
and  28.5 mag arcsec$^{-2}$, respectively, for  reasonable assumptions
of dust extinction.
The present study has demonstrated that metal-poor 
([Fe/H] $\sim$ $-1$) and low surface brightness ($\mu_{B} \sim 30$\,mag\,arcsec$^{-2}$)
stellar halo substructures can be formed along the orbit of the G1 progenitor.
We have also found that the stars in substructures close to the simulated G1
have orbital properties similar to those of  the G1.
Accordingly, future spectroscopic observations 
on radial velocities and metallicities of stars in the G1 clump  
will provide a new clue to the problem  of  
whether the G1 clump can be tidal debris of the G1 progenitor dE,N.

Our simulations suggest that if a dE,N has a higher central dark matter
density (i.e., more compact core),  it cannot be transformed
into a giant globular cluster (G1) because of the survival of its stellar envelope.
Whether a dE,N can be transformed into a globular cluster by
the M31 tidal field (i.e., by galaxy threshing)
depends on whether the M31 tidal force
is stronger than the self-gravitational force of the dark matter
halo of the dE,N {\it at the pericenter} of the orbit of the dE,N.
Therefore we can give {\it the upper limit} of the central dark matter
density of the G1 progenitor dE,N, if we know its pericenter distance from M31. 
The possible central density of the dark matter halo of $\omega$ Cen's progenitor dE,N
is discussed in Bekki \& Chiba (2003), based on the proper motion
data of $\omega$ Cen.

Drinkwater et al. (2003) have recently discovered a new type
of galactic object with  effective radius of $\sim$ 20 pc,
$M_{V}$ $\sim$ $-12$ mag, and  velocity dispersion of $\sim$ 30 km s$^{-1}$
in the Fornax Cluster.
These ultra-compact dwarfs  (UCDs) with $M_{V}$ more than 1 mag
brighter than that of G1 
have been demonstrated to be formed by galaxy threshing in which
dE,Ns  with $M_{V}$ $<$ $-16$ mag
can be transformed into UCDs owing to tidal stripping
of the stellar envelopes of dE,Ns (Bekki et al. 2001, 2003).
We thus 
suggest that UCDs and giant globular clusters such as G1 and $\omega$ Cen
can be regarded as the same class of stellar objects:
The total mass or luminosity  of a   progenitor dE,N is 
the main difference between UCDs and G1.
However, the difference in the location on  
the $M_{\rm V}$-${\sigma}_{0}$ relation between UCD and G1
cannot be explained simply by the galaxy threshing scenario (Bekki et al. 2003).

The present study suggests that if G1 is {\it not} close to  the pericenter  
of its orbit around M31,
the stripped stars from the G1
 progenitor dE,N can be distributed throughout
the M31 outer halo region with  $R_{\rm p}$
$>$ 40 kpc. 
Therefore, future observations on M31's stellar halo 
need to extend  to the current limit of
previous surveys ($R_{\rm p}$ $\sim$ 40 kpc; Ferguson et al. 2002)  to reveal
possible tidal streams and substructures formed from  dE,N. 
Metallicity information of the possible  streams and substructures
in  M31's outer halo  
is also useful to constrain the stellar population of the destroyed G1 progenitor dE,N.
Thus, future deep, high-resolution, wide-area surveys 
of M31's {\it outer} ($R_{\rm p}$ $>$ 40 kpc)
stellar halo by wide-field cameras on large ground-based telescopes (e.g., Suprime-Cam
on Subaru) will enable us to determine whether G1 originates from
an ancient dE,N orbiting M31.  
\begin{acknowledgements}
K.B.  acknowledges the Large Australian Research Council (ARC).
All the simulations described here were performed
with GRAPE 5 systems
at the National Astronomical Observatory in Japan.
%We are  grateful to the anonymous referee for valuable comments,
%which contribute to improve the present paper.
\end{acknowledgements}


\begin{thebibliography}{}

%  \bibitem[1966]{baker} Baker, N. 1966,
%      in Stellar Evolution,
%      ed.\ R. F. Stein,\& A. G. W. Cameron
%      (Plenum, New York) 333


\bibitem[]{}
Bekki, K., Couch, W. J., Drinkwater, M. J.
2001, ApJL, 552, 105

\bibitem[]{}
Bekki, K., Couch, W. J., Drinkwater, M. J., \& Shioya, Y.
2003, MNRAS, 344, 399 

\bibitem[]{}
Bekki, K., \& Chiba, M. 2003, in preparation

\bibitem[]{}
Bekki, K., \& Freeman, K. C. 2003,  accepted in MNRAS

\bibitem[]{}
Binggeli, B., \& Cameron, L. M., 1991, \aap, 252, 27

\bibitem[]{}
Burkert, A. 1994, MNRAS, 266, 877

\bibitem[]{}
C\^ot\'e, P., Marzke, R. O., West, M. J., \&  Minniti, D.
2000, \apj, 533, 869

\bibitem[]{}
Dinescu, D. I. 2002, in ASP Conf. Ser. 265 
Omega Centauri, A Unique Window into Astrophysics.  
ed.  F. van Leeuwen, J. D. Hughes, and G. Piotto (San Francisco: ASP), 608

\bibitem[]{}
Djorgovski, S. G., Gal, R. R., McCarthy, J. K., Cohen, J. G.; de 
Carvalho, R. R., Meylan, G., Bendinelli, O., \&  Parmeggiani, G.
1997, \apj, 474, L19

%\bibitem[]{}
%Drinkwater, M. J., Phillipps, S.,  Jones, J. B., Gregg, M. D.,   Deady, J. H.,
%Davies, J. I., Parker, Q. A., Sadler, E. M., \& Smith, R. M. 2000a,
%\aap, 355, 900

%\bibitem[]{}
%Drinkwater, M. J., Jones, J. B., Gregg, M. D.,  Phillipps, S.  2000b,
%PASA, 17, 227

\bibitem[]{}
Drinkwater, M. J., Gregg, M. D.,  Hilker, M., Bekki, K., Couch, W. J.,
Ferguson, J. B.,  Jones, J. B.,   Phillipps, S. 2003, Nature, 423, 519 

\bibitem[]{}
Durrell, P. R., Harris, W. E., \& Pritchet, C. J. 2000, AJ, 121, 2557


\bibitem[]{}
Ferguson, A. M. N., Irwin, M. J.,  Ibata, R. A., Lewis, G. F., \& Tanvir, N. R
2002, \aj, 124, 1452

\bibitem[]{}
Ferguson, H. C.,  Bingelli, B. 1994, A\&ARv, 6, 67

\bibitem[]{}
Freeman, K. C. 1993, in The globular clusters-galaxy connection,
edited by Graeme H. Smith, and Jean P. Brodie,
ASP conf. ser. 48, p608

\bibitem[]{}
Gebhardt, Karl, Rich, R. M., \&  Ho, L. C. 2002, \apj, 578, L41

\bibitem[]{}
Gnedin, Oleg Y., Zhao, H., Pringle, J. E.,
Fall, S. M.,  Livio, M., \&
Meylan, G. 2002, \apj, 568, L23

\bibitem[]{}
Hilker, M. \&  Richtler, T. 2000, \aap, 362, 895

\bibitem[]{}
Ibata, R. Irwin, M. Lewis, G, Ferguson, A. M. N., \&  Tanvir, N.
2001, \nat, 412, 49

\bibitem[]{}
King, I. R. 1962, AJ, 67, 471

\bibitem[]{}
Meylan, G., Jablonka, P., Djorgovski, S. G., Sarajedini, A., Bridges, T.,
\&  Rich, R. M. 1997, BAAS, 29, 1367

\bibitem[]{}
Meylan, G., Sarajedini, A., Jablonka, P., Djorgovski, S. G.,
Bridges, T., \&  Rich, R. M. 2000, BAAS, 32, 1440

\bibitem[]{}
Meylan, G., Sarajedini, A., Jablonka, P., Djorgovski, S. G.,
Bridges, T., \&  Rich, R. M.  2001, \aj, 122, 830

\bibitem[]{}
Mizutani, A., Chiba, M., \& Sakamoto, T. 2003,  ApJ, 589, L89

\bibitem[]{}
Navarro, J. F., Frenk, C. S.,  \& White, S. D. M.
1996, \apj, 462, 563 

\bibitem[]{}
Rich, R. M., Mighell, K. J., Freedman, W. L., \&
Neill, J. D. 1996, \aj, 111, 768

\bibitem[]{}
Salucci, P.,  Burkert, A. 2000, ApJL, 537, 9

\bibitem[]{}
Sugimoto, D., Chikada, Y., Makino, J., Ito, T., Ebisuzaki, T., \&
Umemura, M. 1990, \nat, 345, 33

\bibitem[]{}
van den Bergh, S.  2000, The Galaxies of the Local Group

\bibitem[]{}
Zhao, H. S. 2002,
in ASP Conf. Ser. 265 
Omega Centauri, A Unique Window into Astrophysics.  
ed.  F. van Leeuwen, J. D. Hughes, and G. Piotto (San Francisco: ASP), 391


\bibitem[]{}
Zinnecker, H., Keable, C. J., Dunlop, J. S., Cannon, R. D., \&  Griffiths,  W. K.
1988, in Grindlay, J. E., Davis Philip A. G., eds, Globular cluster systems in Galaxies,
Dordrecht, Kluwer, p603


\end{thebibliography}
\end{document}